\documentclass[pra, aps,showpacs, notitlepage, groupedaddress,superscriptaddress, twocolumn, numerical]{revtex4-1}

\usepackage{graphicx}
\usepackage{amssymb}
\usepackage{xcolor}
\usepackage{siunitx}
\usepackage{ulem}
\usepackage{nicefrac}
\usepackage{hyperref}

\newcommand{\Gammael}{\Gamma_{\textrm{el}}}
\newcommand{\Gammaloss}{\Gamma_{\textrm{loss}}}

\begin{document}

\title{Controlled doping of a bosonic quantum gas with single neutral atoms}

\author{Daniel Mayer}
\thanks{D.M.~and F.S.~contributed equally to this work.}
\author{Felix Schmidt}
\thanks{D.M.~and F.S.~contributed equally to this work.}
\author{Daniel Adam}
\author{Steve Haupt}
\author{Jennifer Koch}
\author{Tobias Lausch}
\author{Jens Nettersheim}
\author{Quentin Bouton}

\address{Fachbereich Physik und Landesforschungszentrum OPTIMAS, Technische Universit\"at Kaiserslautern, 67663 Kaiserslautern, Germany}

\author{Artur Widera}
\email{widera@physik.uni-kl.de}
\address{Fachbereich Physik und Landesforschungszentrum OPTIMAS, Technische Universit\"at Kaiserslautern, 67663 Kaiserslautern, Germany}
\address{Graduateschool Materials Science in Mainz, Gottlieb-Daimler-Str. 47, 67663 Kaiserslautern, Germany}
\vspace{10pt}

\begin{abstract}
We report on the experimental doping of a $^{87}$Rubidium (Rb) Bose-Einstein condensate (BEC) with individual neutral $^{133}$Cesium (Cs) atoms.
We discuss the experimental tools and procedures to facilitate Cs-Rb interaction.
First, we use degenerate Raman side-band cooling of the impurities to enhance the immersion efficiency for the impurity in the quantum gas. 
We identify the immersed fraction of Cs impurities from the thermalization of Cs atoms upon impinging on a BEC, where elastic collisions lead to a localization of Cs atoms in the Rb cloud.
Second, further enhancement of the immersion probability is obtained by localizing the Cs atoms in a species-selective optical lattice and subsequent transport into the Rb cloud.
Here, impurity-BEC interaction is monitored by position and time resolved three-body loss of Cs impurities immersed into the BEC.
This combination of experimental methods allows for the controlled doping of a BEC with neutral impurity atoms, paving the way to impurity aided probing and coherent impurity-quantum bath interaction.
\end{abstract}

\maketitle

\section{Introduction}
Single impurities coupled to a Bose-Einstein condensate (BEC) form a model system underlying many theoretical proposals to exploit localization of impurities in a quantum gas. The applications of impurity-bath interaction include local probing of various quantum properties of many-body systems \cite{Bruderer2006,Sabin14, Correa2015,Elliott2016,Streif2016,Klein07, Ng08}, or cooling of impurities and quantum information carriers exploiting the properties of quantum fluids \cite{Daley04, Griessner06,Lausch18}. 

An application which has attracted significant attention recently is the simulation of fascinating quantum impurity phenomena in solid state physics. 
For strong interactions, quasi-particles emerge which can have profoundly different properties compared to the uncoupled impurity.
A prominent example is the recently observed formation of strong-coupling polarons in Bose-gases \cite{Hu16,Jorgensen16}, where a detailed understanding of the ground state properties is currently emerging \cite{Bruderer07b, Blinova13, Rath13, Li14,Levinsen2015, Ardilla15,Shchadilova16}. 
More generally, individual impurities with spin degree of freedom can realize an experimental model system for the spin-boson model \cite{Leggett87} or the central spin model \cite{Prokofev00}, describing the properties of a spin in a bosonic or spin bath, respectively.

In order to realize a model system of individual neutral impurities in a quantum fluid, independent control of individual impurities within the BEC is necessary.
This calls for experimental tools facilitating independent positioning of BEC atoms and impurities.
So far, this has been achieved by either exploiting the spin degree of freedom, i.e.~creating spin impurities \cite{Fukuhara2013}, or by using electronically charged impurities such as ions or electrons \cite{Zipkes2010, Schmid2010, Balewski2013}. 
However, these approaches imply either a fixed spin degree of freedom, or further experimental challenges due to the long-range atomic interaction potential for collisions in the quantum gas, which requires much smaller temperatures to enter the $s$-wave interaction regime.
An alternative route is to use neutral impurities, where technically the tools and methods of state preparation and independent position control have to rely on the multi-level structure of impurity and quantum gas. 

This work outlines the experimental toolbox for controlled doping of a $^{87}$Rubidium (Rb) BEC with a small number of individual neutral $^{133}$Cesium (Cs) atoms.
In section~\ref{sec:Methods} we will introduce the methods to prepare a Rb BEC and individual laser cooled Cs atoms in close vicinity, and present two important methods facilitating doping of the BEC: Raman cooling and species selective transport of Cs atoms in an optical lattice.
In section~\ref{sec:Results} we will first demonstrate how the effective Rb-Cs cross section can be enhanced by Cs cooling.
We find that the efficiency to immerse a neutral Cs atom into a Rb gas crucially depends on the question if a single Rb-Cs collision occurs.
Experimentally, we find that the scattered fraction of Cs atoms is increased when lowering the Cs temperature.
Introducing a simple model, we reproduce the qualitative behavior of the Cs immersion probability.
Second, we show how the species-selective optical lattice allows for deterministic immersion of impurities into the BEC by suppressing the axial Cs motion.
Thereby, our work facilitates future studies of controllable impurity-bath interaction. 
\begin{figure*}
	\begin{center}
		\includegraphics[scale=1]{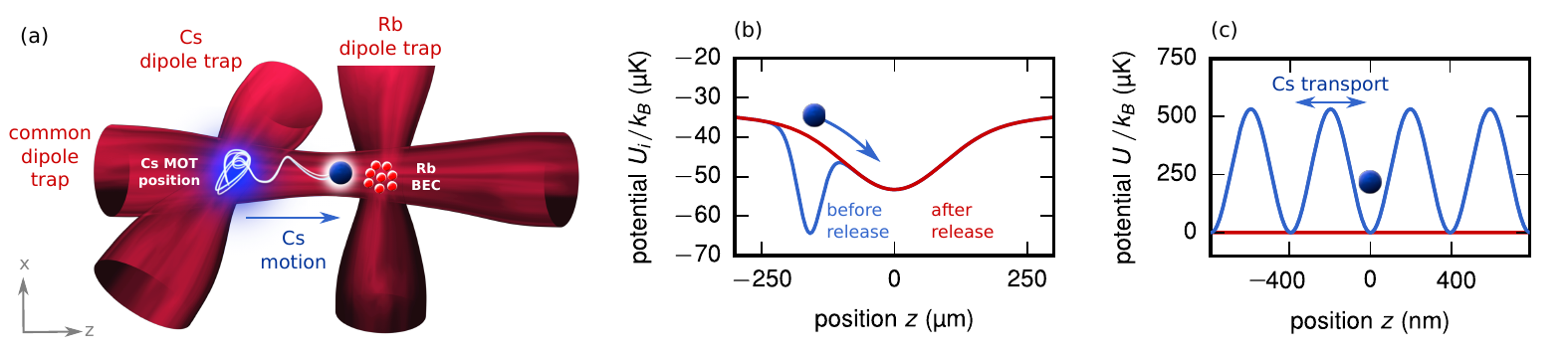}
	\end{center}
	\caption{Experiment overview.
		(a) Sketch of the optical trap configuration.
		(b) Dipole potential for Cs impurities $U_i$ before (blue) and after (red) switching off the Cs dipole trap. 
		The motion of Cs atoms toward the Rb cloud is driven by the potential of the Rb dipole trap.
		(c) A species-selective one-dimensional optical lattice along the $z$ axis creates a repulsive potential for Cs atoms (blue) while at the same time no significant potential is present for Rb (red).
		The lattice is used for spatially resolved fluorescence imaging of Cs atoms. 
		Furthermore, it can be employed to transport Cs atoms along the $z$ direction, by introducing a small detuning between the counter-propagating lattice beams.
	}
	\label{fig:Sketch}
\end{figure*}

\section{Methods} \label{sec:Methods}

Our experiment produces Rb BECs in an all-optical configuration (Fig.~\ref{fig:Sketch}) with typically $10^4$ atoms at temperatures around $300\,$nK at a cycle time of approximately 7\,s, for details see \cite{Hohmann2015}. 
The Rb cloud is trapped in a crossed dipole trap comprising a horizontal beam with waist of $\SI{21}{\micro \meter}$ and a vertical beam of waist $\SI{165}{\micro \meter}$. 
The final trapping frequencies in the combined, cigar-shaped trap are $\omega= 2 \pi \times \SI{700}{\hertz}$  $(2 \pi \times \SI{50}{\hertz})$ in the radial (axial) direction.
The Rb cloud is probed by standard absorption imaging after time-of-flight.

Few Cs atoms are cooled and trapped in a high magnetic field gradient magneto-optical trap (MOT) and subsequently loaded into a crossed dipole trap.
The dipole traps for Cs and Rb share the horizontal trapping beam while the crossing region for the Cs trap is located at a distance of approximately $\SI{170}{\micro \meter}$ from the Rb BEC (see Fig.~\ref{fig:Sketch}). 
Typically, the Cs atoms have temperatures in the order of $\SI{10}{\micro \kelvin}$ after the transfer from the MOT to the dipole trap.

\subsection{Raman cooling}
While ultracold, the temperature of the laser-cooled Cs atoms still is more than one order of magnitude larger than the BEC temperature. The resulting small spatial overlap between the two species turns out to be a central obstacle for controlled impurity immersion. To further reduce the Cs temperature, we employ degenerate Raman-sideband cooling (DRSC), similar to ref.~\cite{PhysRevLett.84.439}. 

\begin{figure*}
	\begin{center}
		\includegraphics[scale=1.]{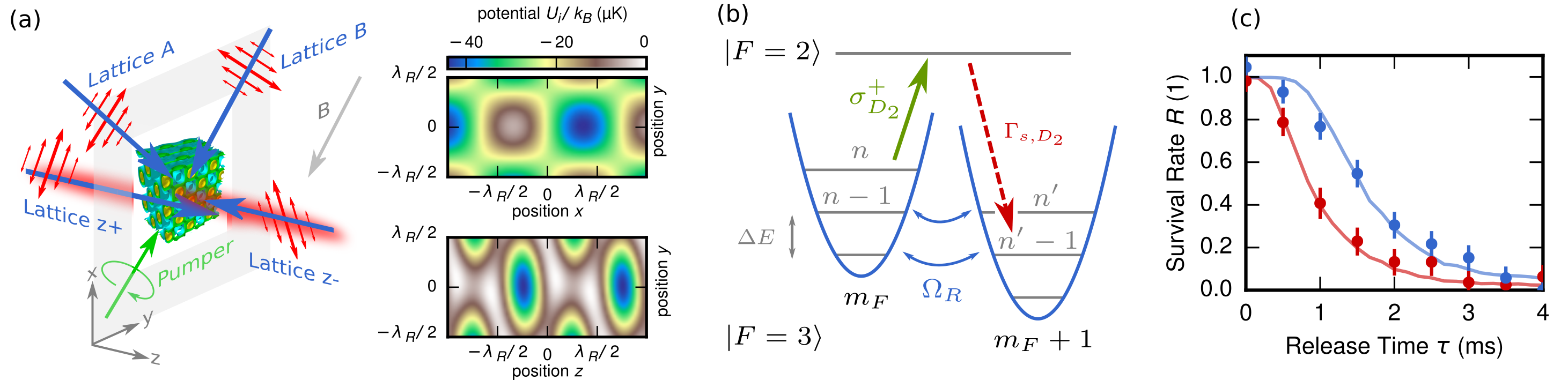}
	\end{center}
	\caption{(a) Raman laser setup. Four Raman beams (blue arrows) intersect at the center of the crossed dipole trap (red shaded area), for details see text. The beams create a 3D intensity distribution, trapping the Cs atoms in the potential wells.
		Two cuts through the 3D potential landscape of the Raman lattice in the $x$-$y$ (top) and $z$-$y$ plane (bottom) at the position of a potential minimum with $\lambda_R = \SI{852}{\nano \meter}$.
		(b) Degenerate Raman sideband cooling scheme (adapted from \cite{PhysRevLett.84.439}).
		(c) In a release-recapture experiment, we measure the temperature of the Cs atoms by comparing the measured survival rate $R$ for different release times (dots) to simulated curves at different temperatures. From the best-fitting simulated curve (lines), we extract a temperature of $\SI{12\pm2}{\micro K}$ (red) when applying an optical molasses pulse to the Cs atoms. After a single Raman cooling pulse, we measure a temperature of $\SI{3.2\pm0.4}{\micro K}$ (blue). 
	}
	\label{fig:RamanSketch}
\end{figure*}

The optical setup combines a near-resonant optical lattice with an optical pumping beam in order to reduce the kinetic energy of the atoms. The optical lattice provides both an optical trapping potential as well as Raman coupling between adjacent $m_F$ states.
The Raman lattice is created by three pairwise orthogonal beams with beam waists of approximately $1.1\,$mm, referred to as A, B and Z+ (Fig.~\ref{fig:RamanSketch}).
The Z+ beam is propagating along the horizontal dipole axis with power of $9\,$mW and is retro-reflected to create a fourth beam Z-.
Beams A and B with a power of $4\,$mW each propagate mutually orthogonal to each other, as well as with respect to the horizontal axis.
They are polarized linearly in the plane defined by their propagation axes, while the beams Z+ and Z- are polarized vertically with a small polarization tilt of roughly $10^\circ$ between the two counter-propagating beams.
The laser frequency of the Raman lattice beams is red detuned to the $\vert F=4\rangle \to \vert F^\prime =4 \rangle$ transition of the Cs $D_2$-line by 6\,MHz, where $F$ is the total atomic angular momentum.
As the atoms are preferably in the $\vert F=3 \rangle$ state during the cooling cycle, this corresponds to a red shift of roughly 9\,GHz.
The interference between the four beams creates a 3D optical lattice pattern with a trap depth around $k_B\times \SI{45}{\micro \kelvin}$ and trapping frequencies of $\omega_\text{lat} = 2\pi \times (70, 29, 28)\,$kHz (with the Boltzmann constant $k_B$).
The lattice beams also drive the Raman transitions in the $F = 3$ manifold with coupling strength $\Omega_R$ and repump Cs atoms from $\vert F=4 \rangle$ into the cooling cycle in the case of off-resonant excitation.
In order to reach degeneracy of vibrational energy levels with quantum number $n$ in adjacent states $\vert m_F + 1, n \rangle$ and $\vert m_F, n-1 \rangle$ in the Raman lattice potential, the magnetic background field $B$ is tuned to an experimentally optimized value of 100\,mG.
The optical pumping beam is resonant on the $D_2$ line and pumps atoms in a $\sigma^+$-transition from the $F = 3$ manifold to the $F^\prime = 2$ manifold where they spontaneously emit a photon ($\Gamma_{s, D_2}$) and decay back to $F = 3$.
The pump beam is right-hand circularly polarized with a small amount of linear polarization and blue detuned by $12\,$MHz to the $\vert F=3\rangle \to \vert F^\prime = 2 \rangle$ transition.
The Lamb-Dicke parameter $\eta = \sqrt{E_R/\hbar \omega_\text{lat} }$ for spontaneous decay is $ \eta = 0.27\,(0.17)$ for the smallest (largest) trap frequency in the Raman lattice, with the single-photon recoil energy $E_R = \hbar^2 k_\text{pump}^2 / (2 m_\text{Cs})$. 
This indicates that the vibrational quantum number is conserved for most of these photon scattering processes, hence the Cs energy is reduced by $\Delta E = \hbar \omega_\text{lat}$ in each Raman cycle and therefore the atoms are cooled.
During the Raman cooling, the position of a Cs atom remains confined in the optical lattice, while the kinetic energy is reduced and the Cs population is pumped to $m_F=3$ with about $75 \, \%$ fidelity.
 
We determine the temperature directly after Raman cooling with a release-recapture technique, where the dipole trapping potential is switched off for a short duration and the remaining fraction of atoms is measured as a function of switch-off durations \cite{Mudrich2002, Reymond2003, Hohmann2016}.
Colder atoms have smaller velocities, therefore leaving the trapping volume more slowly and thus have a higher recapture probability.
We extract the Cs temperature by comparing the experimental data to simulated curves which are generated in a Monte-Carlo simulation for Cs samples at different temperatures.
After a single Raman cooling pulse, we measure a Cs temperature of $\SI{3.2\pm0.4}{\micro \kelvin}$ (Fig.~\ref{fig:RamanSketch}).

\subsection{Independent position control}
While the loading of the two species takes place at distant positions along the shared trapping beam, the atomic density distributions need to be overlapped to facilitate interaction.
A key tool for independent position control of neutral Cs atoms with respect to the Rb cloud is a species-selective lattice (Fig.~\ref{fig:Sketch}). 
Its operation relies on the multi-level structure of atoms. 
Adding all light-shift contributions, for alkali atoms usually the two fine-structure lines are considered, one finds that between the Rb D-lines the resulting dipole potential has a zero crossing at the so-called tune-out wavelength \cite{LeBlanc2007}. 
Consequently, at a wavelength around $790\,$nm, Rb atoms do not experience a dipole potential for a specific choice of light polarization and atomic hyperfine state \cite{Leonard15,Schmidt2016}. 
For Cs atoms, however, the light field exerts a repulsive potential (Fig.~\ref{fig:Sketch}).

Using two counter-propagating laser beams at the tune-out wavelength, we realize a species-selective 1D optical lattice which confines the Cs atoms axially, while the radial confinement originates from the horizontal dipole trap.
A small detuning between the frequencies of the two beams forming the lattice, sets the standing wave into motion and enables controlled transport of Cs atoms from the loading region along the horizontal trapping axis, while Rb remains static at its trap position \cite{Schrader2001}.

\section{Results} \label{sec:Results}
We characterize the immersion of individual Cs atoms into the BEC by two complementary approaches.
In a first step, we employ Raman cooling to effectively compress the spatial distribution of Cs atoms in the trap due to the reduced temperature. 
Thus, the density-density overlap is sufficiently large to study Rb-Cs interaction with individual Cs atoms.
In a scattering experiment of few Cs atoms impinging on a Rb cloud we show that the effective Rb-Cs immersion probability is large enough to thermalize Cs atoms in the Rb BEC via elastic collisions. 
This is indicated by a localization of Cs atoms in the Rb crossed dipole trap. 
Second, we further enhance the immersion efficiency by localizing Cs atoms in the species-selective lattice and transport them to the position of the BEC.
We study the position resolved loss of Cs after being transported to the trap center, which reveals enhanced loss in the vicinity of the Rb BEC. 

\subsection{Effective immersion probability between impurity and cloud}
\label{sec:effective_scattering_crosssection}

\begin{figure*}
	\begin{center}
		\includegraphics[scale=.9]{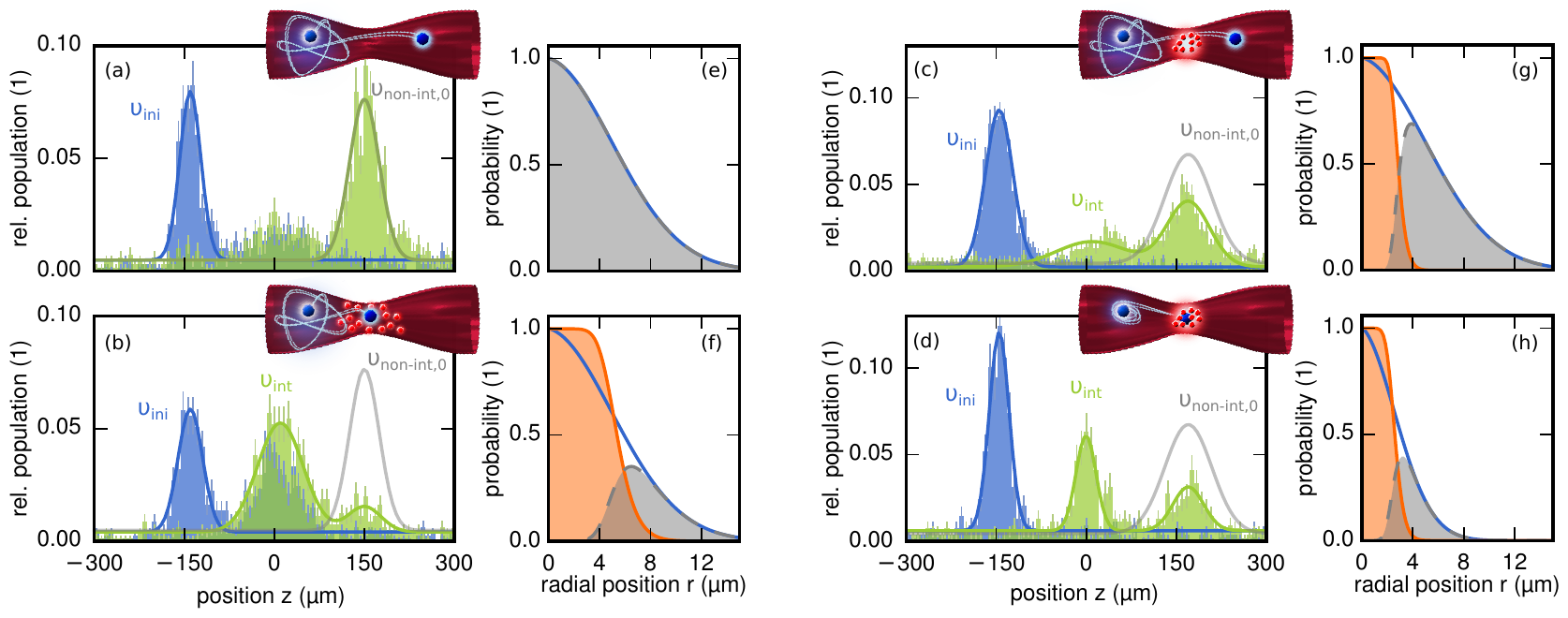}
	\end{center}
	\caption{(a-d) Axial position distribution of individual Cs atoms in the trap for different conditions. 
		Blue, green, and grey are the the initial, and final Cs position distribution after half an oscillation period with and without Rb, respectively.
		From the final distribution, the fraction of interacting ($\nu_{\mathrm{int}}$) and non-interacting Cs atoms ($\nu_{\mathrm{non-int}}$) is extracted.
		The fraction of interacting atoms is calculated as $1 - \nu_{\mathrm{non-int}} / \nu_{\mathrm{non-int, 0}}$, where $\nu_{\mathrm{non-int, 0}}$ is the reference, when no Rb is present.
		(a) Laser-cooled Cs atoms ($T_\mathrm{Cs}\approx 10\,\si{\micro \kelvin}$ oscillate freely, when no Rb is present. 
		(b) When impinging on a thermal Rb cloud ($N_{\mathrm{Rb}}=11 \,\mathrm{k}$, $T_{\mathrm{Rb}}=\SI{1.1}{\micro \kelvin}$), a large Cs fraction ($85\,\%$) undergoes a first collision with Rb and interacts with Rb.
		(c) When impinging on a Rb BEC ($T_\mathrm{Rb}\approx 300\,$nK), the interacting fraction is reduced ($46\,\%$) due a smaller BEC volume. 
		(d) Cs atoms after a series of three Raman cooling pulses ($T_\mathrm{Cs} \approx 3\,\si{\micro \kelvin}$) impinge on a Rb BEC ($N_{\mathrm{Rb}}=23\,\mathrm{k}$, $T_\mathrm{Rb} \approx 300\,$nK). 
		The Raman pulses have effectively compressed the radial probability distribution, leading to an enhanced probability ($74\,\%$) for a first collision.
		(e-h) We use a simple model (see text) to reproduce the qualitative behavior of  the thermalizing fraction: We show radial cuts of the initial Cs density (blue line), the radial scattering probability $P(r)$ (orange shaded line) and the resulting non-interacting fraction (gray shaded dashed line) for parameters comparable to the measurements shown in (a-d). In (g), the non-interacting fraction is increased compared to (f) due to the smaller extent of the Rb cloud. By reducing the Cs temperature with Raman cooling (h), the initial distribution becomes more narrow, resulting in a larger overlap with the scattering area yielding a reduced non-interacting fraction.}
	\label{fig:CsRbScatteringCrossSection}
\end{figure*}

In order to probe the effective Rb-Cs immersion probability, we first prepare a Rb cloud in the Rb crossed trap at various conditions, either in a thermal state with peak densities of $n_0 = \SI{3.6e12}{\centi \meter^{-3}}$ and typical width of $\sigma_r = \SI{2.3}{\micro \meter}$ ($\sigma_z = \SI{31}{\micro \meter}$) radially (axially), or as a BEC with peak densities of $n_0 = \SI{2e14}{\centi \meter^{-3}}$ and Thomas-Fermi radii of $R_r = \SI{1.1}{\micro \meter}$ ($R_z = \SI{15}{\micro \meter}$) radially (axially). 
Cs atoms are trapped in the MOT and subsequently transferred into the Cs crossed trap.

The choice of dipole trap potential depth $- k_B \times \SI{31}{\micro \kelvin}$ ($- k_B \times \SI{53}{\micro \kelvin}$) for Rb (Cs) during the \mbox{Rb-Cs} interaction is a compromise between providing sufficient confinement for reliable Cs atom trapping and avoiding a heating of the BEC at high densities induced by three-body loss.
Due to the strong radial confinement and the similarity of trap frequencies of Cs and Rb, the difference of the gravitational sag is in the order of $\SI{50}{\nano \meter}$ and is negligible compared to typical radial extensions of the atomic clouds ($\geqslant \SI{1}{\micro \meter}$). 

For the measurement, Cs atoms are prepared first directly from the MOT with uncontrolled hyperfine state and temperatures in the order of $T_\mathrm{Cs} \approx \SI{10}{\micro \kelvin}$; or second with an additional Raman cooling step yielding temperatures of $T_\mathrm{Cs}\approx \SI{3}{\micro \kelvin}$ in $m_F=3$.
Subsequently, the Cs atoms are released from the Cs crossed dipole trap and move in the horizontal trapping beam toward the center, where the Rb cloud is prepared (Fig.~\ref{fig:Sketch}). 
After half an oscillation period, the Cs distribution is frozen by quickly applying a strong optical lattice potential, suppressing tunneling as well as thermal diffusion.
After the Rb cloud has been removed from the trap by a resonant laser pulse, the Cs atoms in the 1D optical lattice are detected by in-situ fluorescence imaging. 
The axial Cs density distribution is obtained from hundreds of fluorescence images for each experimental setting.

It has been shown that essentially a single collision of Cs with a Rb bath atom dissipates sufficient energy so that the Cs atom remains trapped in the crossed Rb trapping region \cite{Hohmann2017}.
By contrast, the fraction of atoms which have not experienced a collision will freely oscillate to the turning point of the potential.

Experimentally, half an oscillation period after the release from the MOT, here $17.5\,$ms, the interacting ($\nu_{\mathrm{int}}$) and non-interacting ($\nu_{\mathrm{non-int}}$) Cs fractions are maximally separated and can be well distinguished in our position-resolved fluorescence images (Fig.~\ref{fig:CsRbScatteringCrossSection}(b)). 
The relative number of Cs atoms in the Rb trapping region is thus a measure of the effective immersion probability.
The results of such scattering measurements are shown in fig.~\ref{fig:CsRbScatteringCrossSection} for different situations. 
First, Cs atoms with MOT temperature of $T_\mathrm{Cs}\approx \SI{10}{\micro \kelvin}$ impact a thermal Rb gas with $T_\mathrm{Rb} = \SI{1.1 \pm 0.2}{\micro \kelvin}$ and $11(1) \times 10^3$ atoms.
From the data in Fig.~\ref{fig:CsRbScatteringCrossSection}(b) we conclude that a large fraction of $85\,\%$ of Cs atoms collides with Rb atoms in the cloud and thermalizes with it.
By contrast, if the Rb cloud is condensed ($T_\mathrm{Rb} \approx 300\,$nK), the thermalized fraction is drastically reduced to $46\,\%$ as shown in Fig.~\ref{fig:CsRbScatteringCrossSection}(c).
This is due to the reduced size of the Rb BEC ($\SI{1.1}{\micro \meter}$) compared to the thermal cloud ($\SI{2.4}{\micro \meter}$), and many Cs atoms are radially too far away from the Rb BEC to undergo a first collision. 
Thus, the effective immersion probability is reduced.

In order to model the scattered fraction of Cs atoms, we introduce the position dependent Rb-Cs scattering rate 
$\Gamma(r, z) = \sigma \bar{v} n_b(r, z)$
for a Cs atom at position $(r, z)$. 
Here, $n_b(r, z)$ is the Rb density distribution in the trap including BEC and thermal fraction. Assuming a Maxwell-Boltzmann distributed kinetic energy distribution, the mean relative velocity between Cs and Rb atoms \cite{Mosk2001} is approximated by
\begin{eqnarray}	
\bar{v} = \sqrt{\frac{8 k_b} {\pi} \left( \frac{T_\text{Cs}}{m_\text{Cs}} + \frac{T_\text{Rb}}{m_\text{Rb}}\right)},
\end{eqnarray}	
$\sigma = 4 \pi a^2$ is the scattering cross section with the \mbox{Rb-Cs} \mbox{s-wave} scattering length $a = 645\,a_0$ \cite{Takekoshi2012} and the Bohr radius $a_0$.
Neglecting the radial motion of the Cs atoms in the trap, we assume that they impinge on the Rb cloud in axial direction at velocity $v_z = \sqrt{2E_0 / m_\text{Cs}}$ given by their initial potential energy in the dipole trap $E_0$.
We calculate the scattering probabilities $P(r)$ for these straight trajectories at different radii $r$ from the trap center by evaluating the survival probabilities $P_i$  in $N$ regions of length $ \delta z_i$ along the trajectory in $z$ direction
\begin{eqnarray}
P(r) = 1 - \prod_{i=0}^{N} P_i = 1 - \prod_{i=0}^{N} \exp\left(-\Gamma(r, z_i) \frac{\delta z_i} { v_z } \right).
\end{eqnarray}
Comparing this radial scattering probability to the Cs radial distribution (Fig.~\ref{fig:CsRbScatteringCrossSection} (e-g)), we find that the immersed Cs fraction is reduced for lower Rb cloud temperatures due to the smaller Rb cloud size.
The model also illustrates that the effective immersion probability can be increased again by lowering the temperature of the Cs impurities (Fig.~\ref{fig:CsRbScatteringCrossSection} (h)).

Experimentally, a series of Raman cooling pulses is employed to further cool the Cs atoms. 
We use three Raman cooling pulses of 2\,ms duration each, separated by a waiting time of $\SI{350}{\micro \second}$ between two pulses. 
The waiting time is adjusted to the radial trapping frequency and matches a quarter oscillation period.
The first Raman pulse will hence reduce the Cs atoms' kinetic energy.
After a quarter oscillation period the remaining potential energy is converted to kinetic energy, which is dissipated by the second Raman pulse. 
More than two pulses are necessary because the trap is not perfectly harmonic; the number of three is chosen as a compromise, as every Raman pulse leads to a loss of a small fraction of Cs atoms ($\approx 5\%$).
Fig.~\ref{fig:CsRbScatteringCrossSection}(d) shows the Cs distribution when the Cs atoms have been Raman cooled. 
In this configuration, a fraction of $74\,\%$ of impurities is immersed into the Rb BEC through collisions with the Rb cloud.

This measurements illustrate how the effective immersion probability can been significantly increased, which enables controlled production of individual Cs impurities in a Rb BEC for future investigations.

\subsection{Controlled doping and impurity lifetime}
 
\begin{figure*}
	\begin{center}
		\includegraphics[scale=1]{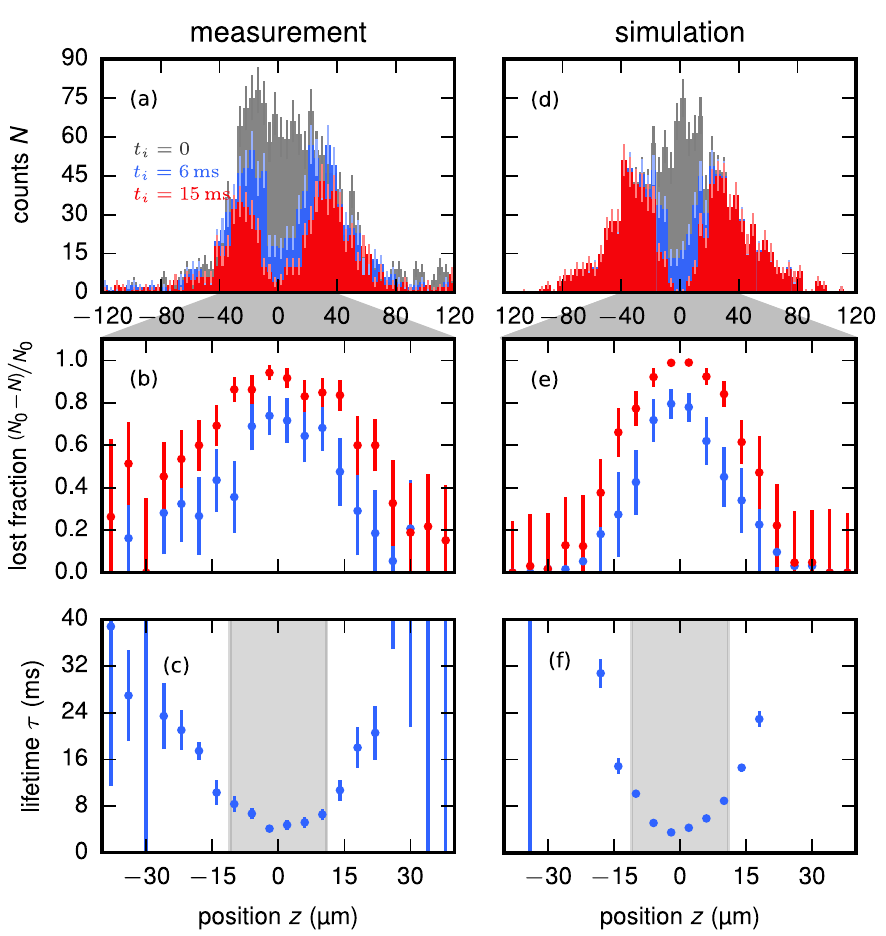}
	\end{center}
	\caption{(a) Axial Cs distribution for different interaction times $t_i$ in the vicinity of the Rb BEC (15\,k atoms, 0.2 condensate fraction).
	(b) The bin-wise lost fraction relative to $t_i=0$ is extracted from these distributions. (c) Assuming an exponential decay, we calculate the position resolved lifetime $\tau$.
	The gray shaded area marks the range of the BEC's Thomas Fermi radius $R_z$.
	We also compare the measurement results to a simulation (d-f) where we simulate the tree-body losses of Cs atoms which are thermalized within a Rb BEC. 
	We find good agreement between experiment and simulation without free parameters.
	The error of the lifetime fit depends first on the initial atom number $N_{k, 0}$ which determines the statistical uncertainties, leading to large errors in regions with few atom counts.
	And second on the resulting value of $\tau_k$ which has to be compared to the maximum experimental interaction duration of $\SI{15}{\milli \second}$.
}
	\label{fig:CsHoleBurning}
\end{figure*}
In a complementary approach to implant Cs impurities in the BEC, we exploit the species-selectivity of our optical lattice to transport Cs impurities into the BEC. 
In addition, after establishing interaction, the lattice is used to pin the Cs position within the Rb cloud.
This allows to resolve the interaction dynamics without motion of the Cs impurities along the horizontal trapping beam. 
We monitor the position resolved loss of Cs atoms for increasing interaction time, which is strongly enhanced in a BEC due to frequent Rb-Rb-Cs three-body recombination events. 

To this end, we prepare a Rb BEC of approximately $15 \times 10^3$ atoms with a Thomas-Fermi radius of $R_z \approx \SI{11}{\micro \meter}$ in axial direction. 
Raman-cooled Cs atoms are loaded into the species-selective lattice and transported into the BEC.
Then, the lattice is used to transport the Cs atoms to various positions around the expected positions of the Rb cloud, intentionally broadening the Cs distribution axially.

Specifically, Cs atoms are distributed by choosing axial transport distances in the range of about \SIrange{145}{195}{\micro \meter}.
The transport lattice depth of $k_B \times \SI{17}{\micro \kelvin}$ ($140\,E_R$) is chosen to yield sufficient confinement for Cs during the transport with an axial trapping frequency of $2\pi \times \SI{80}{\kilo \hertz}$. 
At the same time this choice avoids an excessive heating of the Rb cloud due to off-resonant photon scattering:
For the lattice power used here, the remaining scattering rate on both $D$ lines is $<\SI{13}{\hertz}$.
After the transport, the lattice depth is further reduced (factor of $\approx 1.7$), when fixing the axial Cs position during the Rb-Cs interaction.

Cs atoms experience frequent collisions within the dense Rb cloud. 
In addition to elastic collisions, leading to fast thermalization of the Cs atoms, inelastic Rb-Rb-Cs collisions occur, which result in a loss of Cs atoms through molecule formation. 
The Cs distribution is measured after different interaction times $t_i$, revealing the emergence of a hole at the position of the Rb BEC (Fig.~\ref{fig:CsHoleBurning}\,(a)).
Thus, the losses demonstrate the successful immersion of individual Cs atoms into the Rb BEC.

We extract the position-resolved lifetime of Cs atoms in the BEC by first sorting the Cs positions extracted from the fluorescence images into a histogram with bins of about $\SI{5}{\micro \meter}$ width. 
For each bin $k$ and interaction duration $t_i$, the fraction of atoms lost is calculated as $ P^\text{loss}_k (t_i) = \left(N_{k}(0) - N_k(t_i)\right) / N_{k}(0)$, where $N_{k}(0)$ denotes the initial Cs atom number of bin~$k$.
This normalization allows calculating the position dependent lifetime independently from the shape of the initial Cs distribution (Fig.~\ref{fig:CsHoleBurning}(b)).
For each position bin $k$ the lifetime $\tau_k$ is extracted by fitting an exponential decay $\tilde{P}^\text{loss}_k (t_i) = 1 - \exp(-t_i / \tau_k)$ to the remaining fraction (Fig.~\ref{fig:CsHoleBurning}(c)). 
We find the shortest Cs lifetimes of only few ms at the expected center position of the BEC.

We compare the measurement to a Monte-Carlo model, where thermalization and loss dynamics for independent atomic Cs trajectories $\mathbf{r}_{\textrm{Cs}}$ are simulated, assuming thermalized Cs atoms in the Rb BEC. 
For each atom in the Monte-Carlo sample which is as large as the experimental atom number, the equation of motion $\ddot{\mathbf{r}}_{\textrm{Cs}} = {-\nabla V} / {m_{\textrm{Cs}}}$ in the trapping potential $V$ (dipole trap and gravity) is integrated, as discussed in \cite{Hohmann2017}. 
Elastic collisions with the Rb bath, as well as three-body loss are incorporated by evaluating the probabilities of an elastic collision $P_{\mathrm{el}} = 1 - \exp{ \left(- \Gammael \Delta t \right)}$ and atom loss $P_{\mathrm{loss}} = 1 - \exp{\left( - \Gammaloss \Delta t \right)}$ in each integration step of length $\Delta t$.

For the calculation of elastic rate $\Gammael$ and loss rate $\Gammaloss$, a continuous Rb distribution with the expected in-trap density of the BEC $n_{\textrm{Rb}}(\mathbf{r})$ consisting of thermal background and BEC Thomas-Fermi distribution is assumed.
The elastic collisions rate writes $\Gammael = \sigma \cdot v \cdot n_{\textrm{Rb}}(\mathbf{r}_{\textrm{Cs}})$ with the low-momentum $s$-wave scattering cross section $\sigma$ introduced in section~\ref{sec:effective_scattering_crosssection}.
In contrast to section~\ref{sec:effective_scattering_crosssection}, here the relative velocity $v$ in the collision is evaluated, taking the current velocity $\dot{\mathbf{r}}_{\mathrm{Cs}}$ and the bath temperature into account \cite{Hohmann2017}.
The three-body loss rate writes $\Gammaloss = L_3 \, n_{\textrm{Rb}}^2(\mathbf{r}_{\textrm{Cs}})$ with an independently obtained loss coefficient $L_3 = \SI{28.e-26}{\hertz \, \centi \meter^6}$.
Properties of the BEC were determined independently, so the Monte-Carlo simulation is conducted without free parameters.\\
For the Monte-Carlo simulation, the position resolved lifetime is extracted analogously to the measurement (compare fig.~\ref{fig:CsHoleBurning}(c), (f)) and is in agreement with the measured data. 
The simulation reveals that about $33 \, \%$ of Cs atoms are lost within the Thomas-Fermi radius of the BEC, although only $20 \, \%$ of Rb atoms are condensed, demonstrating the enhanced density overlap of the condensate fraction at the trap center.

While in the previous section it has been shown that interaction of impurities with the BEC through elastic collisions leads to thermalization and a localization of Cs atoms at the trap center, here we show that inelastic collision events can serve as indicator of successful impurity immersion into the BEC.

\section{Conclusion}
We have experimentally studied the interaction of individual Cs impurities with a Rb BEC. 
For laser-cooled impurities from a MOT, we find that the effective Rb-Cs scattering probability is too small to ensure efficient immersion of Cs in the Rb BEC. 
We demonstrate how the scattering probability can then be enhanced by further cooling of the impurity, which we have implemented by a series of Raman cooling pulses.
This effectively compresses the impurity density distribution in the trap. 
We have measured this increased immersion probability by observing thermalization and localization in the Rb cloud.
Further enhancement of the immersion efficiency is obtained by axially localizing the Cs atoms in a species-selective optical lattice and transporting them into the Rb cloud.
A position resolved measurement of the Cs atom loss rate reveals a strong signature of the interaction between Cs impurity and Rb BEC.
Thereby, our results pave the way for future investigations of coherent interactions between impurities and a quantum gas.

\section*{Acknowledgements}
We thank Michael Hohmann and Farina Kindermann for their help in initially constructing the experiment and for initial discussions, and Valeryia Mikhaylova for preparing illustrations of the experimental details.
This work was funded in the early stage by the European Union via ERC Starting grant "QuantumProbe"; equipment was partially contributed by Deutsche Forschungsgemeinschaft via Sonderforschungsbereich (SFB) SFB/TRR185. 
D.M. acknowledges funding  via  SFB/TRR49,  T.L. acknowledges funding by Carl Zeiss Stiftung, and F.S. acknowledges funding by the Studienstiftung des deutschen Volkes.

\bibliographystyle{apsrev4-1}
\bibliography{QSciTech-Biblio}{}

\end{document}